\documentclass{jfm}
\usepackage{graphicx}
\usepackage{epstopdf, epsfig}
\usepackage{amsmath}
\usepackage{float}
\usepackage[dvipsnames]{xcolor}
\usepackage{natbib}
\usepackage{hyperref} 
\usepackage{bm}

\newcommand{\ii}	    {\mathrm{i}}

\newcommand{\dd}	    {{\rm d}}

\newcommand{\beq}		{\begin{equation}}
\newcommand{\eeq}		{\end{equation}}
\shorttitle{Exact planetary waves and jet streams}
\title{Exact planetary waves and jet streams}

\author{ Nick Pizzo\aff{1}\corresp{\email{nicholas.pizzo@uri.edu}} \& Rick Salmon\aff{2}}

\affiliation{\aff{1}Graduate School of Oceanography, University of Rhode Island, Narragansett, RI 02882, USA \aff{2}Scripps Institution of Oceanography, University of California San Diego,
La Jolla, CA 92037, USA }
\begin{document}

\maketitle

\begin{abstract}
We investigate exact nonlinear waves on surfaces locally approximating the rotating sphere for two-dimensional inviscid incompressible flow. 
Our first system corresponds to a $\beta$-plane approximation \textcolor{black}{at the equator} and the second to a $\gamma$ approximation, with the latter describing flow near the poles. 
We find exact wave solutions in the Lagrangian reference frame that cannot be written down in closed form in the Eulerian reference frame.
The wave particle trajectories, contours of potential vorticity and Lagrangian mean velocity take relatively simple forms.
The waves possess a non-trivial Lagrangian mean flow that depends on the amplitude of the waves and on a particle label that characterizes values of constant potential vorticity. 
The mean flow arises due to potential vorticity conservation on fluid particles. 
Solutions over the entire space are generated by assuming that the flow far from the origin is zonal and there is a region of constant potential vorticity between this zonal flow and the waves. 
In the $\gamma$ approximation a class of waves are found which, based on analogous solutions on the plane, we call Ptolemaic vortex waves.
The mean flow of some of these waves resemble polar jet streams. 
Several illustrative solutions are used as initial conditions in the \textit{fully} spherical rotating Navier-Stokes equations, where integration is performed via the numerical scheme presented in \citet{Salmon2023}. 
The potential vorticity contours found from these numerical experiments vary between stable permanent progressive form and fully turbulent flows generated by wave breaking.

\end{abstract}

\begin{keywords}
\end{keywords}

\section{Introduction}

We introduce and analyze two new waves satisfying Euler's equations in Lagrangian coordinates -- one on the $\beta$-plane and another in the $\gamma-$approximation.
These flows are planetary vortex waves and exhibit behavior that is analogous to, but distinct from, motions on the plane, due to variable rotation. 
The \textcolor{black}{variable rotation induces} a Lagrangian mean flow for these waves. 
For particular configurations in the $\gamma-$approximation these waves have a jet-like mean flow and describe polar jet streams. 

Motion on a rotating surface with non-zero curvature is distinct from its planar counterpart due to variations in the planetary vorticity that support wave motions.
These waves represent oscillations in the amount of planetary and relative vorticity about lines of constant latitude. 
\textcolor{black}{The Coriolis parameter describing the planetary vorticity is considered in a hierarchy of approximations by Taylor expanding it about a point of latitude: the lowest order of approximation is the $f$-plane, the first order approximation is the $\beta$-plane and the second order approximation is denoted the $\gamma-$approximation.}
Interestingly, the $\beta-$plane \textcolor{black}{at the equator} corresponds to an embedding space with a metric that has zero curvature.
That is, curvature effects are necessary for a variable Coriolis parameter, but have no direct impact on the geometry of the embedding space.
The $\gamma$-approximation describes motion near the poles of a rotating sphere, \textcolor{black}{where $\beta$ vanishes}.
Solutions to these equations exist on a plane with a variable rotation rate that again supports wave motion.  

A remarkable set of exact solutions to Euler's equations on the plane was found by \citet{Abrashkin1984}.
This class of solutions includes Gerstner's \textcolor{black}{trochoidal} wave, the Kirchoff ellipse, and a class of solutions known as Ptolemaic, or polygonal, vortices, which are a rotationally symmetric analog of Gerstner waves \citep{Guimbard2006}.
For $z=x+\ii y$, where $(x,y)$ are Cartesian coordinates and $\ii^2 =-1$, the \citet{Abrashkin1984} solutions are of the form $z=f(s)e^{\ii \omega_1 t}+g(\overline{s})e^{\ii \omega_2 t}$ where $s=a+\ii b$ for $(a,b)$ Lagrangian particle labels, $\omega_{1,2}$ constants and overline denotes complex conjugate.
The functions $f,g$ are harmonic, but this is the only restriction on their form.
When viewed in a certain reference frame, both Gerstner and Ptolemaic waves may be written as permanent progressive solutions of the form $z=z(a-\omega t,b)$. 

A warning -- exact solutions in one frame generally cannot be written in closed form in a different frame. 
An example of this is the Gerstner wave \citep{LAMB1932}, which has a relatively simple form in the Lagrangian frame but cannot be written in closed form in the Eulerian frame, as this reduces to solving Kepler's equation: 
one would have to find $a$ as a function of $x$ in closed form in the equation $x=a- \sin a$, which is not possible.
Physically, in two dimensions vorticity is conserved along fluid particles and so there is a close connection between the Lagrangian mean flow and the vorticity.
This mass flux Doppler shifts the wave frequency and for certain vorticity distributions the particle trajectories take relatively simple forms. 

On the $\beta-$plane \textcolor{black}{at the equator} and in the $\gamma-$approximation we find a direct analog of Gerstner's wave, but now the frequencies must depend on the particle label $b$ in order to satisfy the vorticity equation.
For finite amplitude waves, a nonlinear correction to the wave phase speed arises, and this can be interpreted as being due to a Lagrangian mean velocity, which leads to a departure in behavior from their planar counterparts.
These waves only exist over a region of the embedding space, as the vorticity contours always form cusps at some critical value of the parameters.
In the plane, this is handled by taking the vorticity to be constant outside of a critical bounding vorticity contour.
For example, Gerstner's wave takes the flow to be irrotational \textcolor{black}{outside of} a bounding vorticity contour that is specified to be the free surface. 
On surfaces approximating the rotating sphere, we take the ambient flow immediately outside of the waves to have constant potential vorticity, and then take the flow to be zonal in the far field.
This yields solutions that are well defined over the entire embedding space. 
 
We consider a range of long time behavior of our exact solutions by integrating them in the fully spherical rotating Navier-Stokes equations, using the scheme developed in \citet{Salmon2023}.
Unlike on the embedding spaces, we now connect the solutions in the $\gamma$ approximation to those on the sphere by choosing the zonal flow in the outer region to be one that exactly solves the equations of motion on the rotating sphere. 
The numerical experiments are a strong test of these solutions. 
We find solutions that range from permanent progressive waves to waves that rapidly overturn and break, generating turbulent motion. 

The plan of the papers is as follows. In Section 2 and Appendix A the equations of motion are presented.
In Section 3 new exact waves are discussed.
These solutions are used as initial conditions for the fully spherical rotating equations of motion in Section 4.
The connections between these flows and jet streams is discussed in Section 5, after which the conclusions of the paper are given. 

\section{$\beta$ and $\gamma$ approximations}
We examine two-dimensional \textcolor{black}{incompressible} flow with coordinates $(x,y)$ obeying conservation of mass \textcolor{black}{(a derivation of these equations from the full equations on a rotating sphere are presented in Appendix A)}
\beq
\frac{\partial u}{\partial x}+\frac{\partial v}{\partial y} = 0,
\label{1}
\eeq
for $(u,v)$ the velocity, and conservation of potential vorticity
\beq
\frac{\partial q}{\partial t}+u\frac{\partial q}{\partial x}+v\frac{\partial q}{\partial y}=\frac{Dq}{Dt}=0,
\label{2}
\eeq
where the potential vorticity is given by
\beq
q=\frac{\partial v}{\partial x}-\frac{\partial u}{\partial y}+F(x,y).
\label{qeul}
\eeq
\textcolor{black}{ and $D/Dt =\partial_t +\bold{u}\cdot\boldsymbol{\nabla}$.}

\textcolor{black}{We consider the dynamics in two locations. First, near the equator $F = \beta y$, with the Coriolis constant being zero at this location, corresponding to a $\beta$-plane approximation. Similarly, we analyze flow near the poles, where $F = f_0-\gamma (x^2+y^2)$ corresponds to the $\gamma$ approximation. Following \citep{phillips1973}, latitude dependent scale factors do not arise in these \textit{classical} expansions in the $\beta$-plane because we are at the equator and in the $\gamma$-approximation because we additionally assume that $\gamma \gg 1$. For a more thorough discussion, see \citet{dellar2011} and the derivation presented in Appendix A.4. Throughout this paper, we repeatedly refer to \textit{the} $\beta-$plane approximation and \textit{the} $\gamma-$approximation, with the understanding that this is one of many such approximations that bears this name. } Here $\beta,\gamma,f_0$ are constants and \textcolor{black}{we take the deformation length scale to be infinite, so that there is no vorticity associated with vertical stretching}.

The main focus of this paper is on solutions in the $\gamma$ approximation. Both of these approximations contain variable rotation and make a simplifying assumption about the curvature of the embedding surface. These approximations require non-zero surface curvature to possess variable rotation rates, but the curvature of the geometry alone is not present in both approximations. 

Consider fluid particles labelled by coordinates $\bold{a}=(a,b)$ and rewrite the Eulerian governing equations in the Lagrangian reference frame $(x(a,b,\tau),y(a,b,\tau))$ where $\tau$ is time and we emphasize that $\partial_{\tau}$ with $(a,b)$ fixed is the time derivative following a fluid particle so that $\partial_{\tau}=D/Dt$. 

\textcolor{black}{We can assign labelling coordinates so that $\dd(mass) = \dd a \dd b$. Defining the reciprocal of the mass-density of the fluid $\rho$ as $J=J(a,b)$, we have } 
\beq
\textcolor{black}{\frac{1}{\rho}\equiv J(a,b) =\frac{\partial (x,y)}{\partial(a,b)}}.
\label{J1}
\eeq
where
\beq
\frac{\partial (x,y)}{\partial(a,b)}\equiv x_ay_b-x_by_a,
\eeq
\textcolor{black}{\citep[for a more thorough discussion of this map see][]{Salmon1998}.}
\textcolor{black}{The labels follow the fluid, from which it follows }
\beq
\frac{\partial J}{\partial \tau} = \frac{\partial(x_{\tau},y)}{\partial (a,b)}+\frac{\partial(x,y_{\tau})}{\partial(a,b)}=\left(\frac{\partial(x_{\tau},y)}{\partial(x,y)}+\frac{\partial(x,y_{\tau})}{\partial(x,y)}\right)\frac{\partial (x,y)}{\partial(a,b)}
\eeq
\beq
=\left(\frac{\partial u}{\partial x}+\frac{\partial v}{\partial y}\right) \frac{\partial (x,y)}{\partial(a,b)}=0,
\eeq
\textcolor{black}{where the term in parentheses vanishes due to the incompressibility of the fluid, i.e. equation \eqref{1}.} Therefore $J(a,b)$ is a time-independent function. Additionally, we require that $J(a,b)$ does not change sign \textcolor{black}{so that the mass-density $\rho$ stays finite}.

Equation \eqref{2} requires the potential vorticity to be conserved on fluid particles. \textcolor{black}{The relative vorticity in the Lagrangian frame is}
\beq
v_x-u_y=\frac{\partial (v,y)}{\partial(x,y)}+\frac{\partial (u,\textcolor{black}{x})}{\partial(x,y)}=\frac{\partial (a,b)}{\partial(x,y)}\left(\frac{\partial (x_{\tau},x)}{\partial(a,b)}+\frac{\partial (y_{\tau},y)}{\partial(a,b)}\right)
\eeq
so that \eqref{qeul} becomes
\beq
q(a,b)J\textcolor{black}{(a,b)} = \frac{\partial(x_{\tau},x)}{\partial(a,b)}+\frac{\partial(y_{\tau},y)}{\partial(a,b)}+J\textcolor{black}{(a,b)} F(x,y).
\label{q1}
\eeq
Our task is to solve \eqref{J1} and \eqref{q1} subject to the constraint that $J(a,b)$ and $q(a,b)$ are time independent and that $J(a,b)$ does not change sign. 

Although the \textcolor{black}{fluid acceleration} is simpler in the Lagrangian frame, the above analysis shows that the nonlinear mapping between frames can turn linear operators, such as a partial derivative, into nonlinear operators. This leads to coupled partial differential equations that are different in form from their Eulerian counterparts and therefore we must employ different strategies to find classes of exact solutions. Although the Lagrangian reference frame is considered here, there is the possibility of finding exact solutions in a frame that is neither purely Eulerian nor Lagrangian, but a mixture of the two (see the formulation given by \citealt{Virasoro1981}). \\

To better understand flow on the $\beta-$plane and in the $\gamma$ approximation in the Lagrangian reference frame, we present zonal flow solutions to these equations. These flows prove useful for patching together solutions over the entire embedding space, as is discussed in \S 3.3. In the $\beta$-plane, zonal flow takes the form 
\beq
x= a +U(b)\tau; \quad y = b.
\eeq
Here $J(a,b)=1$ and $q(a,b) =U'+\beta b$, where $U$ is the unspecified mean flow and a prime represents a derivative with respect to $b$. Contours of constant potential vorticity are horizontal lines.  

In the $\gamma-$approximation, zonal flow takes the form
\beq
x = A_0\sin (ka-(\omega+\Omega(b))\tau )e^{kb}; \quad y = A_0\cos (ka-(\omega+\Omega(b))\tau ) e^{kb},
\eeq
where $A_0,\omega$ and $k$ are constants. These particle locations imply 

\beq
J =A_0^2k^2 e^{2kb} \quad q =f_0-\frac{\gamma}{k^2} J+2(\omega+\Omega)+\frac{\Omega'}{k}.
\eeq
The velocity is entirely in the azimuthal direction and contours of constant potential vorticity take the form of circles. Zonal flow solutions for the rotating sphere are presented in Appendix A. 

\textcolor{black}{We can also write down Rossby wave solutions in the Lagrangian frame for the $\beta-$plane approximation, }
\beq
\textcolor{black}{x=a; \quad y = b+\epsilon \cos (ka-\omega \tau),}
\eeq
\textcolor{black}{where $(\epsilon, k,\omega)$ are yet to be specified constant. These expansions imply $J(a,b)=1$, while }
\beq
\textcolor{black}{q(a,b) = \beta (b+\epsilon \cos (ka-\omega \tau))+\epsilon \omega k \cos (k a-\omega \tau). }
\eeq
\textcolor{black}{For $q(a,b)$ to be $\tau-$independent, we require }
\beq
\textcolor{black}{\omega = -\frac{\beta}{k},}
\eeq
\textcolor{black}{in agreement with the classical result for the Rossby wave dispersion relationship in the Eulerian frame.}

\textcolor{black}{In the $\gamma-$ approximation, the situation is complicated by the geometry of the domain. In the Eulerian frame, \citet{leblond1964} shows that the vorticity contours have radial dependence in the form of Bessel functions, while the dispersion relationship is a function of roots of these Bessel functions. As far as we can tell, there is no simple Lagrangian analog here, an illustration that there are exact solutions in the Eulerian frame that cannot be simply written in the Lagrangian frame. The situation is even more complicated for Rossby-Haurwitz waves on the sphere. For both the $\gamma$-approximation and the full spherical solutions, the waves fill the domain, in strict contrast to the waves found here, which are spatially compact. }

\section{Exact solutions}

We now present a description of waves on the $\beta$-plane and in the $\gamma-$approximation. They cannot cover the entire embedding space as the function $J(a,b)$ would then change sign and the vorticity contours would form cusps. Instead, we write down the particle locations of these waves over some limited region of the embedding space. We then describe ambient conditions outside of these areas, where we specify the flow to have constant potential vorticity over a limited region, before being smoothly connected with a region of zonal flow. We consider each of these regions of the flow in turn. 
 
\subsection{Waves on the $\beta-$plane}
Our strategy is to take known planar solutions from \citet{Abrashkin1984} and to allow the frequency to have label dependence. We first look for waves in the $\beta-$plane to build up intuition on how to tackle the more complicated $\gamma$ approximation. The analog of Gerstner waves on the $\beta-$plane is given by
\beq
x= a+ U(b)\tau - \epsilon \sin \theta e^{kb}, \quad y=b+\epsilon \cos \theta e^{kb},
\eeq
where 
\beq
\theta = ka - ( \omega-k U(b) ) \tau
\eeq
and the Lagrangian mean flow $U$ is specified by
\beq
U(b)=\frac{\beta}{2k^2}\left(\epsilon^2k^2 \frac{e^{2kb}}{2}-k b\right),
\label{Ugerst}
\eeq
$\epsilon$ is the amplitude of the wave and $k,\omega$ the (constant) wavenumber and frequency and $b\le 0$. We note that the mean velocity goes to $\infty$ as $b\to- \infty$, so that these flows, analgous to constant shear flow in an infinite fluid in the plane, are not physical over the entire embedding space. 
These waves imply
\beq
J = 1-\epsilon^2k^2 e^{2kb}
\eeq
and
\beq
qJ = 2\omega \epsilon^2k^2 e^{2kb}+\beta \left(\frac{1}{2k}-\epsilon^4 k^3 e^{4kb}+b\right). 
\eeq
$J$ does not change sign as long as $\epsilon k< 1$, which also corresponds to lines of constant potential vorticity remaining single valued and differentiable. In the limit that $\beta=0$, we return to the Gerstner wave solution in the plane. The particle trajectories for these waves are no longer perfect circles as they are in the plane, and instead are trochoids \textcolor{black}{with a distinct shape from the trochoidal contours of constant vorticity}.  \textcolor{black}{Note, in the plane gravity acts as a restoring force for the waves which oscillate vertically and attenuate in amplitude away from the free surface. On the $\beta-$plane, planetary vorticity is the restoring force, with the waves oscillating about lines of fixed latitude and decaying far away from the equator. }

For $b\ll 0$, $J\to 1, U\to -\beta b/2k$ and $q \to \beta (1/2k +b)$ so that $U$ and $q$ are linear in $b$ and easily related to one another far below the limiting contour $(b=0)$. The curvature $U''=\beta \epsilon^2 k^2 e^{2kb} $ is largest near the surface and then rapidly attenuates with depth as the bottom row in figure 1 shows. The geometry of lines of constant Lagrangian mean flow and vorticity are shown in figure 1 for $k=2$ and $\epsilon =1/2$. 

\textcolor{black}{The dispersion relationship, equation \eqref{Ugerst}, is distinct from Rossby waves as it has $b$ dependence. For $\epsilon$ small, $U\approx -\beta b/2k$, so that these waves have an inverse wavenumber dependence, like the Rossby wave and also propagate to the left when $b<0$. We can then interpret these waves as being incident on a steady shear flow, analogous to the situation for Gerstner waves, which propagate on a steady Eulerian current that is equal and opposite to the Stokes drift. }




\begin{figure}
    \centering
    \includegraphics[width = .75\textwidth]{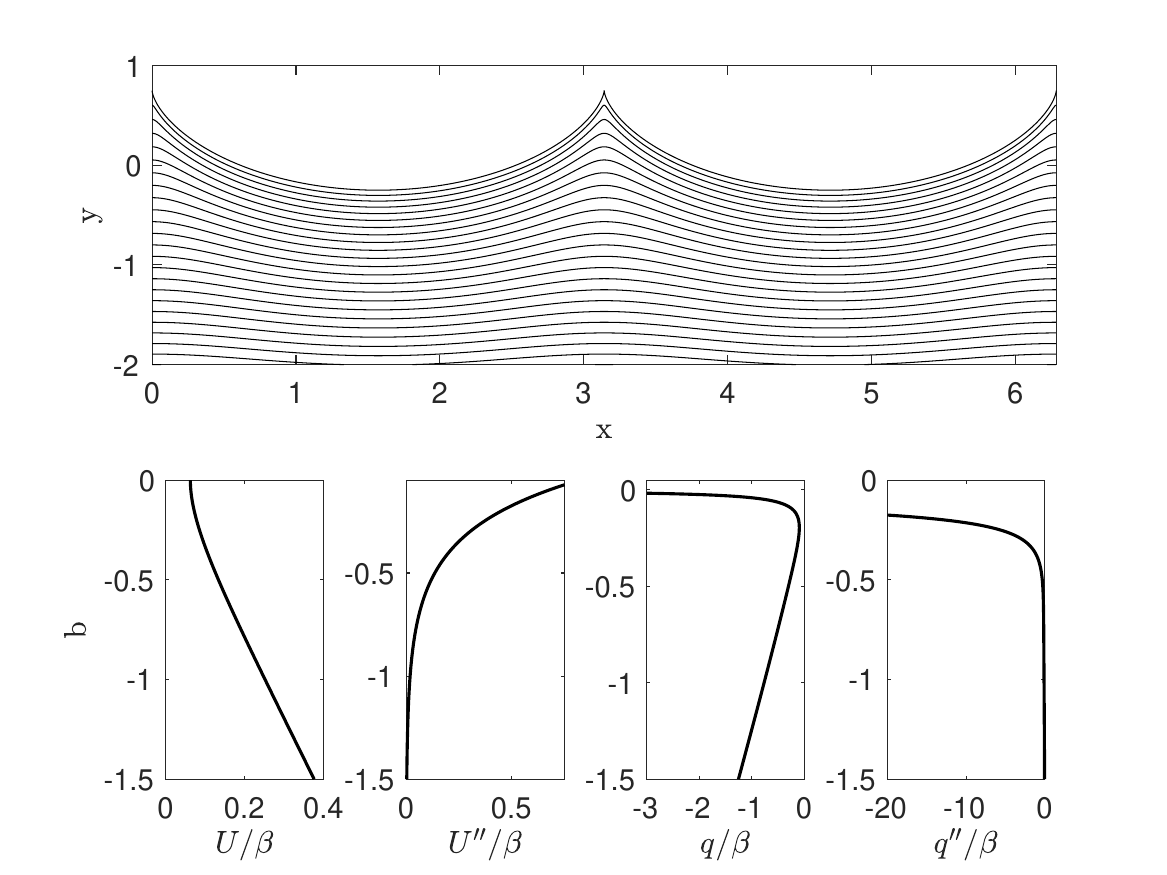}
    \caption{Top: geometry of lines of constant vorticity and Lagrangian mean velocity. Here, $k=2, \epsilon =1/2, b\le 0$ and $\omega=0$. Second row, left: the Lagrangian mean velocity for these waves as a function of $b$ and its curvature. Second row, right: The vorticity and its curvature. Both curvature terms rapidly go to zero for $b\ll 0$. }
    \label{fig:experiment}
\end{figure}

\subsection{Ptolemaic vortex waves in the $\gamma-$approximation}

The main focus of this paper is on solutions in the $\gamma-$approximation near the poles, corresponding to $F=f_0  - \gamma (x^2+y^2) $, \textcolor{black}{where in our non-dimensionalized coordinates the radius of the sphere we are approximating is 1 (see Appendix A), so that $f_0 =\gamma$ }. Our strategy is to once again start from a planar solution and take the rotation rate to depend on a particle label. Planetary Ptolemaic vortex waves, with the nomenclature based on analogous solutions in the plane \citep{Abrashkin1984, Abrashkin1996, Guimbard2006} and the fact that the particle trajectories can be epitrochoids \textcolor{black}{(i.e. the curve generated by tracing a point along a radial line of a circle rolling along another circle)}, are described by
\beq
x/A_0=e^{kb}\sin \theta -\epsilon e^{nk b}\sin n\theta,
\label{xptol}
\eeq
\beq
y/A_0=e^{kb}\cos\theta + \epsilon e^{nkb}\cos n\theta,
\label{yptol}
\eeq
where 
\beq
\theta = ka-(\omega+\Omega(b)) \tau,
\eeq
for $k,\omega, A_0$ constants.
We define $\tilde{x}=x/A_0$ and $\tilde{y}=y/A_0$ and drop the tildes from hereinafter. 

If 
\beq
\Omega(b)= \frac{\gamma }{2n} \left(e^{2kb}-\epsilon^2 n e^{2k nb}\right),
\label{Uptol}
\eeq
then 
\beq
J = k^2\left(e^{2kb}-\epsilon^2 n^2 e^{2nkb}\right)
\eeq
and 
\beq
qJ = \frac{k^2 }{n}\left(\gamma (-2+n)e^{4kb}+\gamma(1+2n)\epsilon^4 n^3 e^{4knb}\right.-
\eeq
\beq
\left.(f_0+2\omega)ne^{2knb}-(f_0-2n\omega)n^3\epsilon^2 e^{2knb}\right).
\label{ptol_pv}
\eeq
The term $\Omega$ should be interpreted as follows: define $\varphi$ implicitly as
\beq
\tan \varphi = \frac{y}{x}
\eeq
then 
\beq
\langle \varphi_{\tau}\rangle=\frac{1-n}{2}(\omega+\Omega),
\eeq
for $\langle \rangle\equiv (2\pi)^{-1}\int_0^{2\pi} \dd a$, so that $\Omega$ is related to the mean azimuthal angular rate of change. The zonal velocity corresponding to this, at this level of approximation, is given by $r\varphi_{\tau}$ where $r^2=x^2+y^2$. Therefore, even if $\Omega$ is relatively weak, for planetary flows $r$ when put in dimensional form can be large implying that these mean flows need not be weak. Depending on the sign of $n$ and $b$ this velocity does not necessarily diverge over the region in the embedding space that the waves are defined over, unlike the scenario in the $\beta-$plane.

\begin{figure}
    \centering
    \includegraphics[width = .6\textwidth]{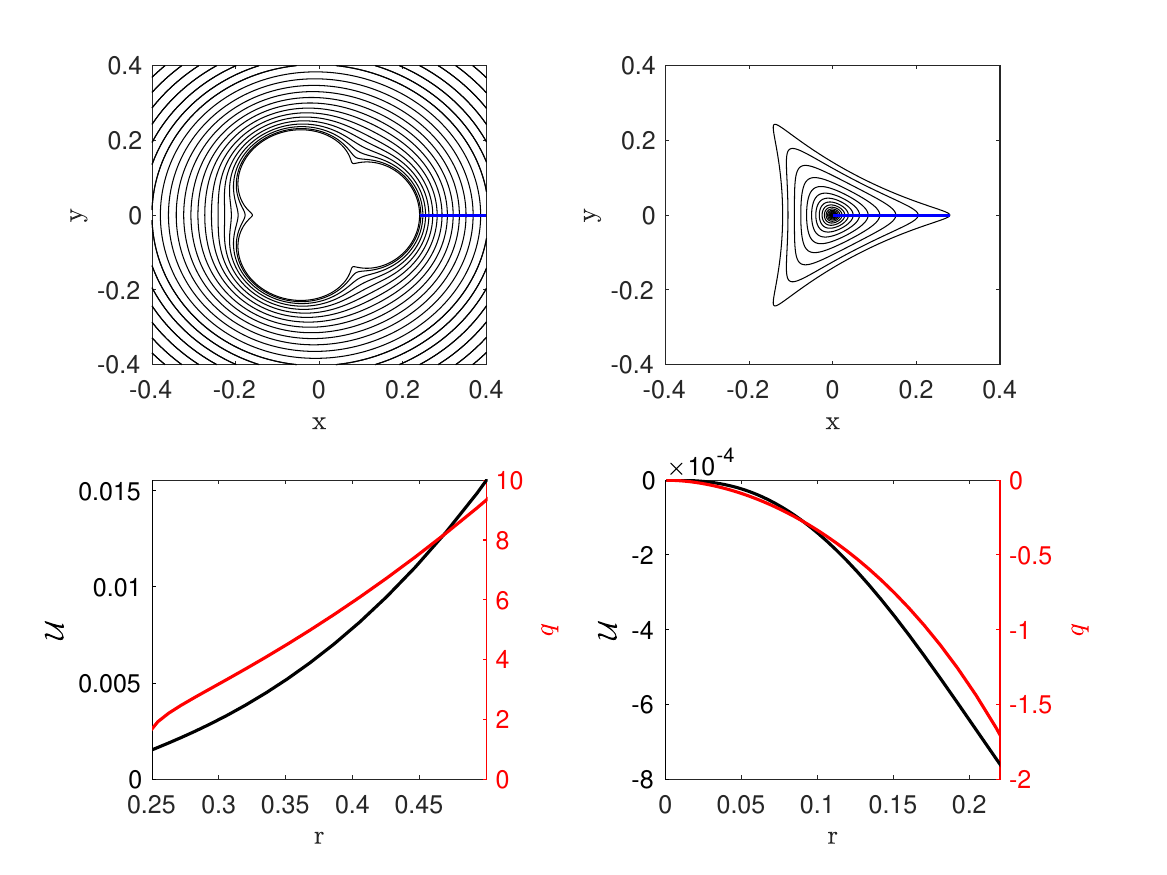}
    \caption{Two Ptolemaic vortex waves. The top row shows the geometry of lines of constant potential vorticity and Lagrangian zonal mean flow. The blue line segments show the transect where $\mathcal{U}$ (in black, with values given by the left ordinate) and $q$ (in red, with values given by the right ordinate) are plotted in the bottom row. The left column shows a wave with $n=-4, A_0=0.2, \epsilon =0.2, \gamma = 1,\omega=0$ and $b\ge 0$. The zonal Lagrangian mean flow and potential vorticity, as a function of $r=\sqrt{x^2+y^2}$, both increase away from the boundary of the wave. Right column: a wave with $n=2, A_0=0.2, \epsilon =0.4, \gamma =1,\omega =0$ and $b\le 0$. The potential vorticity and Lagrangian zonal mean flow decrease in magnitude towards the boundary of the wave.  }
    \label{fig:experiment}
\end{figure}

The mean zonal velocity is 
\beq
\mathcal{U}\equiv \langle r\varphi_{\tau}\rangle =
\eeq
\beq
\frac{1}{2\pi}\int_0^{2\pi} \frac{e^{kb}\left(e^{2k(1-n)b}-n\epsilon^2 +(1-n)\epsilon e^{k(1-n)b} \cos ((1+n)\theta )\right)(\omega + \Omega)}{\sqrt{e^{2k(1-n)b}+\epsilon^2 +2\epsilon e^{k(1-n)b} \cos ((1+n)\theta)  }}\dd a
\eeq
which may be integrated to give 
\beq
\mathcal{U} =
\eeq
\beq
\frac{ e^{knb}}{\pi}\left((1-n)(e^{(1-n)kb}+\epsilon)E(m)+(1+n)(e^{(1-n)kb}-\epsilon) K(m)\right) (\omega+\Omega),
\eeq
where $K,E$ are the complete elliptical integral of the first and second kind, respectively, and 
\beq
m = \frac{4 \epsilon e^{(1-n)kb}}{(e^{(1-n)kb}+\epsilon)^2}.
\eeq

When $n>0$ we take the label $b$ to be negative so that the waves are described for $b\in(-\infty,0)$ and $r\in(0, A_0(1+\epsilon))$. When $n<0$, we take $b\in (0,\infty)$ and $r\in(A_0(1-\epsilon),\infty)$.

\begin{figure}
    \centering
    \includegraphics[width = 1\textwidth]{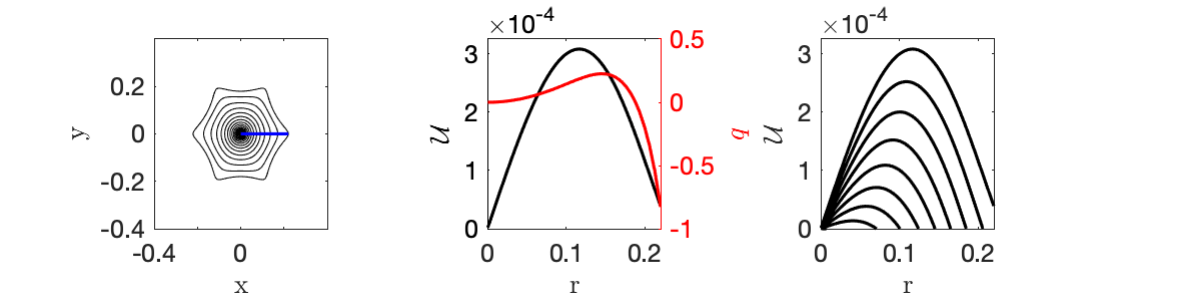}
    \caption{Ptolemaic wave with  $n=5, \epsilon =0.1, A_0=0.2, \gamma =1$ and $b\le 0$. Geometry of contours of constant Lagrangian mean flow and potential vorticity. The blue line is where the mean flow and potential vorticity are computed in the next two panels.  Center: $\omega=0.004$ is non-zero here yielding a localized maximum in the zonal Lagrangian mean flow and potential vorticity, corresponding to jet stream motion near the edge of this vortex wave. Right: the Lagrangian mean flow for $\omega$ varying between $\pm 0.016$ in equal increments, showing that the location of the maximum Lagrangian mean velocity varies as a function of $\omega$.  }
    \label{fig:experiment}
\end{figure}

The parameters $n$ and $\epsilon$ must be chosen to keep $J$ same signed and the contours of $q$ single-valued. These contours are of the form (at $b=t=0$)
\beq
x=\sin \theta_0 -\epsilon \sin n\theta_0,
\eeq
\beq
y=\cos\theta_0 + \epsilon \cos n\theta_0,
\eeq
for $\theta_0 = ka$ and have a cusp when $x_a=y_a=0$, or 
\beq
k\cos \theta_0 -\epsilon nk \cos n\theta_0=-k\sin \theta_0 -\epsilon nk \sin n\theta_0=0.
\eeq
A solution to this is $\theta_0=0$ and $\epsilon  n=-1$ so that we must have 
\beq
|\epsilon n|< 1.
\eeq

Figure 2 shows two Ptolemaic vortex waves. The top row shows the geometry of lines of constant zonal Lagrangian mean flow $\mathcal{U}$ and potential vorticity $q$. The first wave has $n=-4$ and $b\ge 0$. The  right column displays $n=2$ with $b\le 0$. The blue line segments represent the location where $q$ and $\mathcal{U}$ are computed in the bottom row. Both waves have $\omega=0$ and potential vorticity and zonal mean Lagrangian flows that change monotonically. 

In figure 3 we take $n=5$, $b\le 0$ and choose $\omega$ to be non-zero. The mean flow and potential vorticity are shown in the second panel. Both contain a critical point. The mean flow is peaked and represents a jet-like flow. The mean flow for various values of $\omega$ is shown in the right column. Increasing values of $\omega$ have critical points at increasing values of $r$.


The nonlinear term in $\Omega$ arises due to the variable Coriolis parameter. The amplitude squared dependence of the frequency corresponds to the existence of a four wave resonance (\textcolor{black}{c.f. triad resonance for Rossby waves}) and has stability implications for these waves (see \citealt{Pizzo2023} for analogous considerations of surface gravity waves).
Linear stability analysis for these waves is pursued elsewhere -- instead we generate solutions that may be tested numerically -- a much stronger check on the long time existence of these waves. 

\subsection{Ambient conditions and far field behavior}
The waves presented above cannot exist everywhere when $\epsilon$ is non-zero, as the function $J$ would change sign and the potential vorticity contours would become multi-valued. Therefore they exist over some range of label space $(a,b)$. This is analogous to the scenario described by \citet{Abrashkin1984} and discussed in more detail in \citet{Salmon2020}. We extend the discussion presented in \citet{Salmon2020} to generate solutions that are valid over the entire $\beta$-plane and $\gamma-$plane by prescribing ambient conditions outside of the waves. 

In the plane the flow may be taken to have uniform vorticity outside of the bounding curve given by $b=$ constant. \textcolor{black}{A simple analog is to assume that our planetary waves are surrounded by regions of uniform potential vorticity, which may be subsequently connected to far-field regions that have zonal flow. Physically, regions of constant potential vorticity can occur from wave breaking that leads to potential vorticity mixing and homogenization where the vorticity contours have most curvature, that is, near their bounding contours.}


In both scenarios we must specify regions where the potential vorticity $\nabla^2 \psi +f_0-\gamma(x^2+y^2) $ is constant, where $\psi$ is the stream function associated with the \textcolor{black}{relative vorticity of the flow} in this yet-unspecified region \textcolor{black}{evaluated near the equator (the $\beta$-approximation considered in this paper) or near the poles (the $\gamma-$ approximation considered in this paper)}. The potential vorticity can be discontinuous across this boundary, but the velocity must remain continuous. This sets the boundary condition on $\psi$: the tangent velocity of the particle locations provided in \S 3 should be equal to $\psi_{\tilde{n}}$, for $\tilde{n}$ the normal to the contour. $\psi$ is then determined at each point in time and specifies a solution everywhere valid on the $\beta-$ and $\gamma$-plane. 
Of central importance here is the fact that we need not specify the actual trajectories in this region, which can be complicated because of the geometry, as knowledge of $q$ is enough to specify the problem once the continuity of the velocity field has been ensured. 

The region over which the potential vorticity is uniform must be finite for the relative vorticity to stay bounded. Subsequently, when $n>0$ we can connect the constant potential vorticity region to a region of the zonal flow solutions presented in \S 2. The choice of $\omega$ and $\Omega$ for the zonal flow solutions again allows the velocity field to remain continuous. Figure 4 shows a sketch of these scenarios, depending on the sign of n in the $\gamma$ approximation. For $n<0$, the polar cap is taken to have constant potential vorticity and outside of this the flow takes the form of a Ptolemaic wave. When $n>0$, the inner region is the Ptolemaic wave, while immediately outside of this wave the flow has constant potential vorticity and in the far field the flow is zonal. The first two panels are on the $\gamma-$plane while the right panel shows the configuration on southern hemisphere. For this scenario we take the far field to be zonal flow solutions on the fully rotating sphere (see equation \ref{zonal_sphere}).  


\begin{figure}
    \centering
    \includegraphics[width = 1\textwidth]{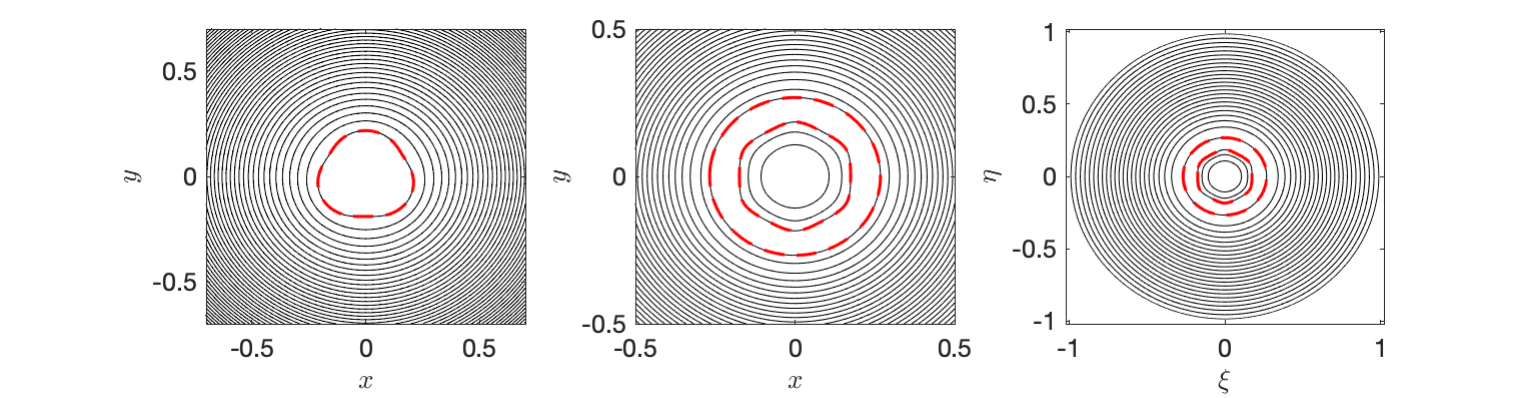}
    \caption{Black: contours of potential vorticity and Lagrangian mean flow. The first two panels show the $\gamma$-approximation and the third are stereographic coordinates on the southern hemisphere of the sphere. The leftmost panel shows the case for $n=-2$. The red dashed line partitions the two regions: inside the polar cap the solution has uniform potential vorticity while beyond that the potential vorticity is given by equation \eqref{ptol_pv}. As the radius of the contours gets large, the Ptolemaic solutions tend towards zonal flow. The central panel shows a Ptolemaic solution with $n=5$. There are three regions here: the innermost region has potential vorticity given by \eqref{ptol_pv}; the region between the two dashed red lines has uniform potential vorticity while the outer region is zonal flow in the $\gamma$-approximation. Right panel: the same as the center panel, but now we take the flow in the outer region to describe zonal flow on the rotating sphere, as given by \eqref{zonal_sphere}. The entire southern hemisphere is shown in stereographic coordinates $(\xi,\eta)$.  }
    \label{fig:experiment}
\end{figure}

\section{Numerics}

\citet{Salmon2023} presented a numerical model of two-dimensional flow on the sphere using stereographic coordinates and a generalization of Arakawa's method developed by \citet{Salmon1989}. This solver can be run inviscidly or viscously and provides an opportunity to examine the solutions presented above. The numerical formulation is Eulerian in nature and so we must carefully provide the initial potential vorticity conditions in this frame. In addition to mass and potential vorticity conservation, the Gauss constraint requires the potential vorticity to integrate to zero over the sphere. 

By construction Arakawa's method conserves discrete analogues of energy, potential vorticity and enstrophy when the viscosity vanishes. The equations of motion are solved in two sets of stereographic coordinates in the unit disc, one corresponding to the northern hemisphere and the other to the southern hemisphere. At the equator we require the solutions to be continuous. These discs are covered in quadrilateral elements and the nodal values of the stream function and potential vorticity serve as the dependent variables of the model. 

Viscosity on non-Euclidean surfaces has been a source of confusion as various reasonable sounding constraints lead to different forms of the final form of the viscosity. Following the recommendation of \citet{gilbert2014}, we use a viscosity that conserves angular momentum, which takes a particularly simple form in stereographic coordinates. There is an error in the form of the viscosity presented in \citet{Salmon2023}, which is corrected in \citet{salmon2024}.   

\begin{figure}
    \centering
    \includegraphics[width = 1\textwidth]{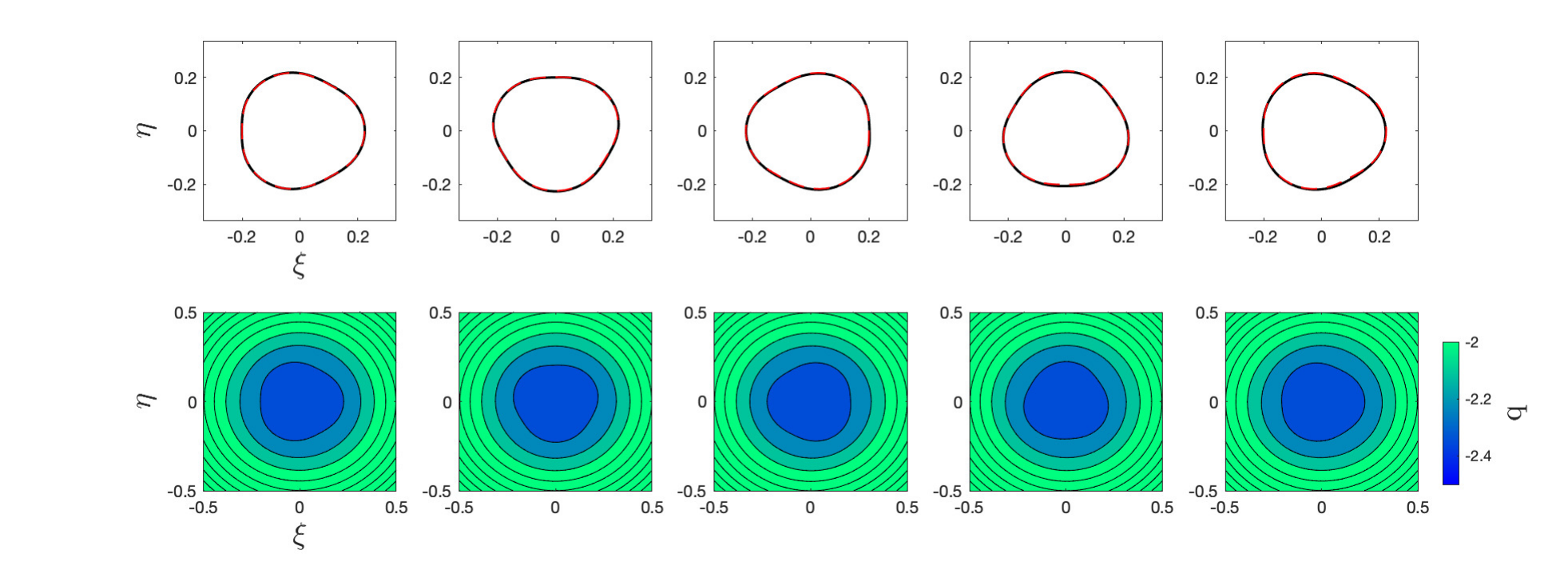}
    \caption{Contours of constant potential vorticity for a Ptolematic wave with $\epsilon =0.075, a_0=0.2, n=-4$ and $\gamma = 10$ integrated over a period of revolution. The region in the polar cap has constant potential vorticity. The top panel shows the flow of one contour in black, versus the theoretical prediction given by the dashed red line. The bottom shows the evolution in the southern hemisphere. Note, $(\xi,\eta)$ are the stereographic coordinates as defined in Appendix A. }
    \label{fig:experiment}
\end{figure}

We integrate the waves presented in \S 3. 
As discussed in the previous section, for $n<0$, we require a constant potential vorticity region at the polar cap surrounded by the Ptolemaic waves. For $n>0$ we prescribe a region of constant potential vorticity outside of the wave region. Then, a third region of potential vorticity, using the known exact zonal flow solutions on the sphere, is chosen so that the velocity remains continuous and the vorticity tends towards zero as the equator is approached. 

In all scenarios we mirror the potential vorticity distribution in the northern hemisphere with that in the southern hemisphere to automatically satisfy the Gauss constraint. We choose $\omega$ so that the potential vorticity exactly vanishes at the equator. This allows us to avoid introducing a vortex sheet at the equator. The initial conditions are chosen to illustrate the range of behavior that these waves exhibit, but we do not perform a thorough examination of parameter space. \textcolor{black}{ We start by illustrating a solution where $n<0$ that has stable vorticity contours for the relatively small value of $|n\epsilon|$ considered. Then, we consider a solution where $n>0$ but $|n\epsilon|$ is taken to be large, so that the waves rapidly overturn and break. Finally, we examine vorticity contours which again have permanent progressive form but in this instance take $n>0$. }

First, in figure 5, we show the evolution of waves with $\epsilon = 0.075, A_0=0.2, n=-4$ and $\gamma = 10$. The top panel shows the evolution of the contour at $b=0$, in black, and the theoretical prediction in red. The two curves are in agreement. Note, these curves are shown in the stereographic plane for the fully spherical geometry. The evolution of the potential vorticity in the southern hemisphere is shown in the bottom panel. 

In figure 6, we show the evolution of a wave with $\epsilon =0.075, A_0=0.2, n=2$ and $\gamma = 10$. The evolution of the potential vorticity is shown. These waves are unstable and \textit{break} almost immediately. The potential vorticity contours become multivalued and the flow near the polar vortex becomes turbulent and potential vorticity from higher latitudes is advected to lower latitudes.

\begin{figure}
    \centering
    \includegraphics[width = 1\textwidth]{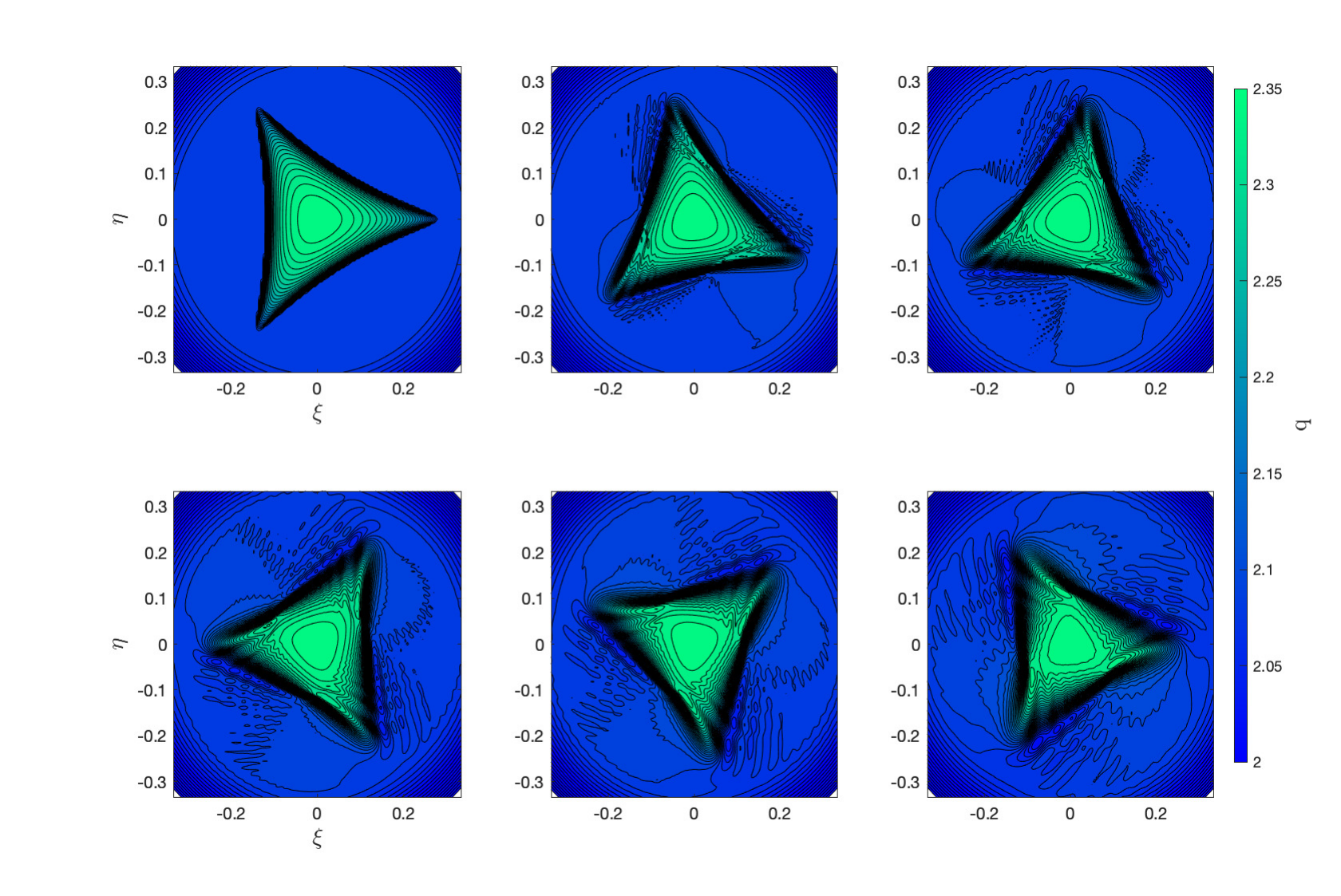}
    \caption{Contours of potential vorticity for Ptolematic waves with $\epsilon =0.4, A_0=0.2, n=2$ and $\gamma = 10$. The region outside of the waves starts with uniform potential vorticity, while the flow in the far field is zonal. These waves rapidly go unstable and overturn and break, generating turbulent flow and advecting high potential vorticity into the region outside of the waves, showing that mass is being transported by these breaking events.}
    \label{fig:experiment}
\end{figure}

Finally, in figure 7 we consider a wave with $\epsilon =0.05, A_0=0.2, n=5, \gamma = 10, \omega = 0.004$, following the configuration shown in figure 3. The geometry of the solutions is reminiscent of the polar jet stream on north pole of Saturn, as is further discussed in the conclusion. The flow retains its structure over the integration time, which exceeds a wave period. \textcolor{black}{Viewed in an inertial frame, the waves are rotating.}

\begin{figure}
    \centering
    \includegraphics[width = 1\textwidth]{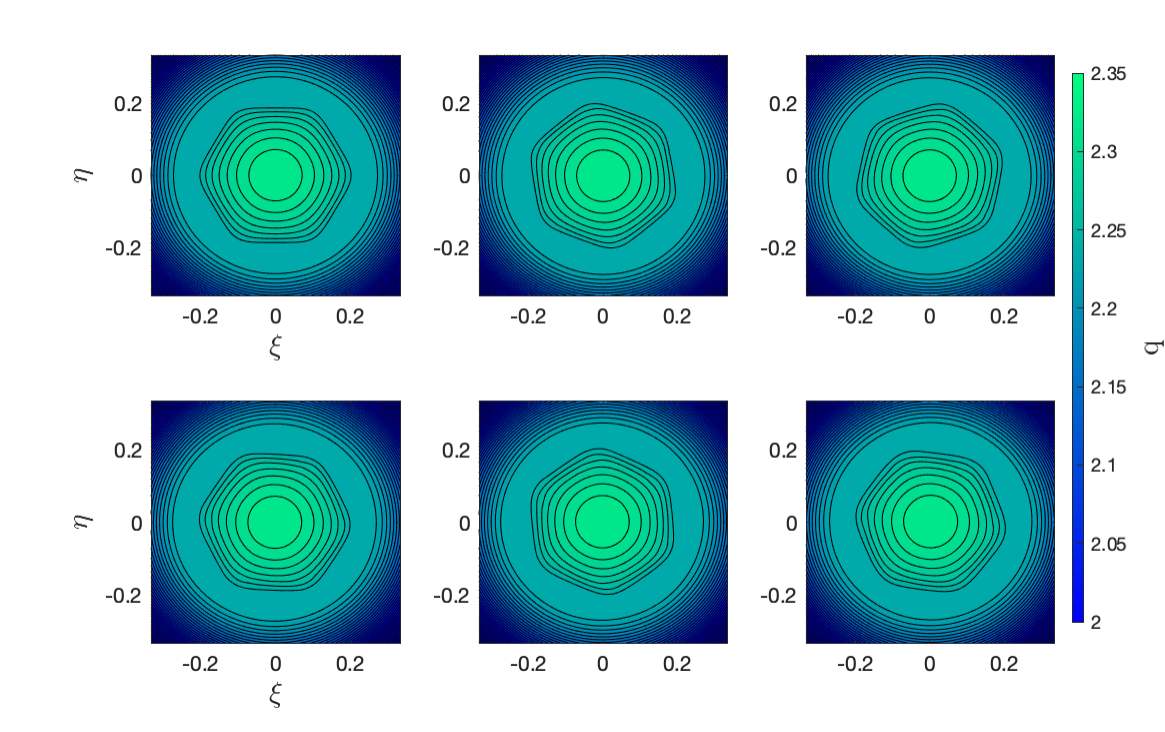}
    \caption{Contours of potential vorticity for Ptolematic waves with $\epsilon =0.05, A_0=0.2, n=5$ and $\gamma = 10$. The flow outside of the waves has uniform potential vorticity, before becoming zonal in the far field. The Lagrangian mean flow corresponds to the distribution drawn in figure 3, with a jet like structure. The configuration is approximately stable over the integration time. The geometry of the solutions is reminiscent of the polar jet stream on north pole of Saturn, as is further discussed in the conclusion. }
    \label{fig:experiment}
\end{figure}

\section{Discussion}
The generation of the vortex waves considered in this paper, and their degree of symmetry, arise as an accident of the initial conditions.
Their geometry, stability and potential vorticity distributions in the numerical simulations invites comparison between these flows and polar jet streams on Earth and other planets such as Saturn.
The meandering of the jet stream is still an active area of research \citep{nakamura2018} particularly under a changing climate \citep{Hoskins2015}.
The flows studied here are two-dimensional and unforced, so any comparison with the real atmosphere is admittedly qualitative. 

The Voyager spacecraft first observed the hexagonal flow pattern on Saturn's north pole \citep{Godfrey1988}.
Later, the Cassini mission \citep{ingersoll2020} captured the hexagon in more detail \citep{sayanagi2017}.
These observations show that the hexagon pattern is quasi-stationary \textcolor{black}{in a frame rotating with the atmosphere}, even though the jet velocity is approximately 125 $ms^{-1}$.
The pattern has been observed for more than three decades, implying a long time stability.
Additionally, the relative vorticity gradient is extremely strong across the jet, i.e. it is a relative vorticity front (see the criteria discussed in \S 12.7.3 of \citealt{sayanagi2017}).
\textcolor{black}{The hexagonal solutions presented in figure 7 are qualitatively consistent with these criteria but a quantitative comparison is left for future work.} \\


Exact solutions have been proposed on the sphere in an Eulerian frame \citep{Crowdy2003, Crowdy2004} that require singular vorticity distributions in the form of point vortices.
The $\gamma$ approximation was employed by \citet{Siegelman2022} to explain the vortex crystal structure found on the pole of Jupiter.
The waves in this manuscript are chosen over parameter regimes so that they avoid singular behavior in the form of potential vorticity contour cusps or multi-valued solutions. 

We have presented two new waves in Euler's equations on the $\beta$-plane \textcolor{black}{at the equator} and in the $\gamma$ approximation.
They are motivated by analogous motions on the plane, but differ crucially in that they are describing wave motion where the restoring force is planetary vorticity.
Solutions are constructed with prescribed potential vorticity distributions outside of the wave fields.
Several examples are considered, but our analysis was not a thorough examination of the broad range of solutions that can be constructed. 
Additionally, only two dimensional flows have been considered here -- analogous exact planar solutions can be unstable to three dimensional high frequency perturbations \citep{leblanc2004, Guimbard2006}. \\

Acknowledgements: We thank the anonymous referees for their helpful comments which have led to a significantly improved manuscript. N.P. thanks Professors Jonathan Wurtele and Robert Littlejohn for hosting him in the Department of Physics at UC Berkeley, where part of this work was completed. \\

Declaration of interests. The authors report no conflict of interest.

\appendix

\section{Derivation of governing equations}
We now derive the equations presented in \S 2. 

\subsection{General equations on the sphere in stereographic coordinates}
Consider two-dimensional incompressible flow on the unit sphere, where
\beq
X^2+Y^2+Z^2 = 1 
\eeq
for $(X,Y,Z)$ the Cartesian coordinates. We use the stereographic coordinates $(\xi,\eta)$ to describe these motions. To derive the governing equations on a surface $(X,Y,Z)=(X(\xi,\eta),Y(\xi,\eta),Z(\xi,\eta))$, begin by taking the square of a line element 
\beq
\dd s^2 =\dd X_i \dd X_i =\frac{\partial X_i}{\partial \xi_j} \frac{\partial X_i}{\partial \xi_k} \dd \xi_j \dd \xi_k=g_{jk}\dd \xi_j \dd \xi_k,
\eeq
where $(\xi_1,\xi_2)=(\xi,\eta)$, $g_{jk}$ is the metric tensor and Einstein summation is assumed throughout. When the coordinates $(\xi,\eta)$ are orthogonal, the metric tensor is diagonal and we have 
\beq
\dd s^2 = h_1^2 \dd \xi^2 + h_2^2 \dd \eta^2 
\label{ds}
\eeq
where
\beq 
g_{11}\equiv h_1^2 = \left(\frac{\partial X}{\partial \xi}\right)^2+\left(\frac{\partial Y}{\partial \xi}\right)^2+\left(\frac{\partial Z}{\partial \xi}\right)^2
\eeq
and
\beq 
g_{22}\equiv h_2^2 = \left(\frac{\partial X}{\partial \eta}\right)^2+\left(\frac{\partial Y}{\partial \eta}\right)^2+\left(\frac{\partial Z}{\partial \eta}\right)^2
\eeq
are the tensor components. The kinetic energy of a particle on this surface can be found by dividing equation (\ref{ds}) by $2(\dd t)^2$, so that 
\beq
\frac{1}{2}\left(\frac{\dd s}{\dd t}\right)^2 \equiv \frac{1}{2}(U^2+V^2) = \frac{1}{2}\left(h_1^2\left(\frac{\dd \xi}{\dd t}\right)^2+h_2^2\left(\frac{\dd \eta}{\dd t} \right )^2 \right)
\eeq
where the velocities are given by 
\beq
U = h_1\frac{\dd \xi}{\dd t}; \quad V= h_2\frac{\dd \eta}{\dd t}.
\label{vel}
\eeq
The stereographic coordinates are \citep[see, e.g., ][]{Needham1997}
\beq
\xi =\frac{X}{1-Z}, \quad \eta = \frac{Y}{1-Z},
\label{xieta}
\eeq
with inverse mapping 
\beq
X=\frac{2\xi}{1+\xi^2+\eta^2}, \quad Y=\frac{2\eta}{1+\xi^2+\eta^2}, \quad Z = \frac{-1+\xi^2+\eta^2}{1+\xi^2+\eta^2},
\label{xyz}
\eeq
and the metric components of this conformal map are
\beq
h_1=h_2 \equiv h = \frac{2}{1+\xi^2+\eta^2}.
\eeq
Next, a unit of area of the fluid is 
\beq
dA = h^2 \dd \xi \dd \eta
\label{dA}
\eeq
so that the kinetic energy T is
\beq
T = \int \frac{1}{2}(U^2+V^2) h^2 \dd \xi \dd \eta
\eeq
\beq
 = \int \frac{1}{2}(U^2+V^2) J \dd a \dd b
\eeq
in the Lagrangian reference frame, where $(a,b)$ are particle labels and we define
\beq
J \equiv h^2 \frac{\partial (\xi,\eta)}{\partial (a,b)}.
\label{J}
\eeq


The time independence of $J$ can be established directly by assuming that the area of the fluid does not change. As the fluid has constant density, this implies conservation of mass. Equation (\ref{dA}) implies
\beq
h'^2 \dd \xi' \dd \eta' = h^2\dd \xi \dd \eta
\label{infmass}
\eeq
where 
\beq
\xi' = \xi(a,b) +\delta \tau \xi_{\tau}(a,b), \quad\eta' = \eta(a,b) +\delta \tau \eta_{\tau}(a,b)
\eeq
and $h' =h(\xi',\eta')$. Equation (\ref{infmass}) is equivalent to 
\beq
h'^2 \frac{\partial (\xi',
\eta')}{\partial (a,b)} = h^2 \frac{\partial (\xi,
\eta)}{\partial (a,b)}
\label{infmass2}
\eeq
so that to first order in $\delta \tau$, equation (\ref{infmass2}) implies
\beq
\frac{\partial (h^2\xi_{\tau},\eta )}{\partial (a,b)}-\frac{\partial (h^2\eta_{\tau},\xi )}{\partial (a,b)} = 0,
\eeq
from which it follows that
\beq 
\frac{\partial J}{\partial \tau} = 0,
\eeq
with $J$ defined in equation (\ref{J}). For the mapping to be invertible $J$ must not change sign. 


To find the vorticity, begin with the circulation $\Gamma$ of the fluid
\beq
\Gamma = \oint \frac{\dd X_i}{\dd t} dX_ i, 
\eeq
where the contour is moving with the fluid on our surface. Using the definition of the fluid velocity and a Green's identity we have
\beq
\Gamma = \int hU \dd \xi +hV \dd \eta
\eeq
\beq
=\iint \left(\frac{\partial (hV)}{\partial\xi}-\frac{\partial (hU)}{\partial\eta}\right) \dd \xi \dd \eta
\eeq
\beq
=\iint h^2 q \ \dd \xi\dd \eta,
\eeq
where
\beq
q =\frac{1}{h^2}\left(\frac{\partial}{\partial \xi} \left(h^2\eta_{t} \right)-\frac{\partial}{\partial \eta}\left(h^2\xi_{t}\right)\right).
\eeq
In Lagrangian coordinates this becomes
\beq
q = \frac{1}{h^2}\left(\frac{\partial(h^2\xi_{\tau},\xi)}{\partial(\xi,\eta)}+\frac{\partial(h^2\eta_{\tau},\eta)}{\partial(\xi,\eta)} \right) = \frac{1}{h^2}\frac{\partial(a,b)}{\partial(\xi,\eta)} \left(\frac{\partial(h^2\xi_{\tau},\xi)}{\partial(a,b)}+\frac{\partial(h^2\eta_{\tau},\eta)}{\partial(a,b)} \right)
\eeq
which can be rewritten as
\beq
qJ = \frac{\partial(h^2\xi_{\tau},\xi)}{\partial(a,b)}+\frac{\partial(h^2\eta_{\tau},\eta)}{\partial(a,b)}
\label{q0}
\eeq
where $q=q(a,b)$, which follows from Kelvin's circulation theorem and implies the vorticity is conserved along fluid particles. When $h=1$, the mass and vorticity constraints reduce to those found on the plane. 

We rewrite the system in a reference frame rotating at angular velocity $\Omega$ about the z-axis. This rotation does not change the mass conservation equation, but the potential vorticity becomes (see \citealt{Salmon2023} equation 34)
\beq
q\to q+2\Omega z = q+ 2\Omega \frac{-1+\xi^2+\eta^2}{1+\xi^2+\eta^2}
\label{q}
\eeq
where in Lagrangian coordinates $q$ is defined by equation (\ref{q0}). \\

Zonal flow may also be written down for this system, and takes the form 
\beq
\xi = A_0\sin (ka-(\omega+\Omega(b))\tau )e^{kb}; \quad \eta = A_0\cos (ka-(\omega+\Omega(b))\tau ) e^{kb},
\eeq
where 
\beq
J= k^2 \text{sech}^2 (kb),
\eeq
\beq
qJ=2\varsigma \text{tanh} (kb) +k \text{sech}^2 (kb) (-2k \text{tanh} (kb) (\omega+\Omega)+\Omega').
\label{zonal_sphere}
\eeq
Just like in the $\beta$ and $\gamma$ approximations, we may choose $\omega$ and $\Omega(b)$ such that the velocity has a prescribed value at the boundary and the vorticity contours take the form of zonal flow. \textcolor{black}{We have utterly failed at finding more interesting exact solutions on the sphere in Lagrangian coordinates.}

\subsection{Approximate governing equations}
We may perform the asymptotics at the order of the Lagrangian, so that the conservation laws are readily available, or to the final system of equations presented in the previous section. We choose the later for clarity of presentation. There are two assumptions used to simplify the system (see the related discussion in \citealt{phillips1973} and also the informative overview in \citealt{dellar2011}). First, we assume that disturbances are small deviations from some rest position. That is, 
\beq
\xi = \xi_0+\epsilon \tilde{\xi} (a,b,t); \quad \eta = \eta_0+\epsilon \tilde{\eta} (a,b,t)
\eeq
for $\epsilon \ll 1$ and $(\xi_0,\eta_0)$ constants.

Our second condition is that $\Omega$ is large and we write 
\beq
\Omega = \frac{\omega}{\epsilon}
\eeq
\textcolor{black}{which yields the classical form of these approximations without additional terms \citep{dellar2011}. Note, the non-dimensional rotation rate $\Omega$ for a Rossby number of $0.02$ and a characteristic velocity of 10 $m/s$  is about 50 on Earth -- see the discussion in \citet{Salmon2023}. }
\subsection{$\beta-$plane}
Take $\xi_0$ and $\eta_0$ to be non-zero so that equations (\ref{J}) and (\ref{q0}), with q defined by (\ref{q}), become  
\beq
J =\epsilon^2 h_0^2[\tilde{\xi},\tilde{\eta}] +O(\epsilon^3)
\eeq
and
\beq
qJ = \epsilon^2 h_0^2 \left([\tilde{\xi}_{\tau},\tilde{\xi}]+[\tilde{\eta}_{\tau},\tilde{\eta}]\right)+
\eeq
\beq
8\epsilon h_0^4[\tilde{\xi},\tilde{\eta}]   \omega \left(1-(\eta_0^2+\xi_0^2)^2+4\epsilon (-2+\xi_0^2+\eta_0^2)(\eta_0\tilde{\eta_1}+\xi_0 \tilde{\xi}) \right)+O(\epsilon^3)
\eeq
where 
\beq
h_0 = \frac{2}{1+\xi_0^2+\eta_0^2}
\eeq
and
\beq
[\xi,\eta]\equiv \frac{\partial(\xi,\eta)}{\partial(a,b)}.
\eeq
To illustrate the structure of the equations, first let $\xi_0=0$ and $\eta_0=1$, corresponding to flow near the equator which yields the following equations 
\beq
J=\epsilon^2[\tilde{\xi},\tilde{\eta}]
\eeq
and
\beq
qJ = \epsilon^2 \left([\tilde{\xi}_{\tau},\tilde{\xi}]+[\tilde{\eta}_{\tau},\tilde{\eta}]\right)+2 \omega \tilde{\eta}  J. 
\eeq
Let $\epsilon^2 \hat{J}=J$ and drop the hats for clarity of presentation. Defining \textcolor{black}{(where we recall the radius of the sphere is 1, which must be taken into account to establish the equivalence of the units of this relationship)}
\beq
2\omega \equiv \beta,
\eeq
the governing equations become 
\beq
J = [\xi,\eta],
\label{jbeta}
\eeq
and
\beq
qJ = [\xi_{\tau},\xi]+[\eta_{\tau},\eta] +\beta J \eta. 
\label{qjbeta}
\eeq
which describe our $\beta-$plane approximation near the equator. When $\beta=0$ we return the equations governing motion in the plane \citep{Abrashkin1984, Salmon2020}. 

Next, let $\xi_0=0$ and keep $\eta_0$ arbitrary. The equations of motion become 
\beq
J=h_0^2 \epsilon^2 [\tilde{\xi},\tilde{\eta}]
\eeq
\beq
qJ = \epsilon^2 h_0^2\left([\tilde{\xi}_{\tau},\tilde{\xi}]+[\tilde{\eta}_{\tau},\tilde{\eta}]\right)+2\Omega J  (\mu + \epsilon \lambda \tilde{\eta}) ,
\eeq
where 
\beq
\mu =  h_0^2(1-h_0), \quad \lambda = -4h_0^2\eta_0(-2+\eta_0^2),
\eeq
and
\beq
h_0=\frac{2}{1+\eta_0^2}.
\eeq
Defining $\epsilon^2 \hat{J} =J$ and again letting $2\omega = \beta$ the governing equations become (dropping the hats again)
\beq
J = [\xi,\eta],
\eeq
\beq
qJ = [\xi_{\tau},\xi]+[\eta_{\tau},\eta] +\mu\beta J + \beta\lambda  \eta J. 
\eeq
Redefining $q$ as $q-\mu\beta$ removes the first term from the right hand side of the vorticity equation while $\lambda$ can be absorbed into the definition of $\beta$, leaving a system of equations that is mathematically equivalent to (\ref{jbeta}) and (\ref{qjbeta}).

\subsection{$\gamma$-approximation}
To find our approximate equations of motion near the (south) pole, let $\xi_0=\eta_0=0$ and $\Omega =\omega/\epsilon^2$ so that 
\beq
J = 4\epsilon^2 [\tilde{\xi},\tilde{\eta}]+O(\epsilon^3),
\eeq
\beq
qJ = 8\omega  [\tilde{\xi},\tilde{\eta}]+ 4\epsilon^2 \left([\tilde{\xi}_{\tau},\tilde{\xi}]+[\tilde{\eta}_{\tau},\tilde{\eta}]\right) -8 \omega J (\tilde{\xi}^2+\tilde{\eta}^2)+O(\epsilon^3).
\eeq
Let $\epsilon^2 \hat{J}= J, \hat{\xi}=2\tilde{\xi}, \hat{\eta}=2\tilde{\eta}$ and define
\beq
2\omega = \gamma
\eeq
so that our governing equations become, dropping the hats,
\beq
J = [\xi,\eta],
\label{jgamma}
\eeq
\beq
qJ = f_0 J+ [\xi_{\tau},\xi]+[\eta_{\tau},\eta] -\gamma J (\xi^2+\eta^2)
\label{qjgama}
\eeq
as we concluded from mapping the Eulerian formulation presented in \S 2. \textcolor{black}{Note, in the text we take $f_0=\gamma/\epsilon^2$, which maybe reabsorbed into the definition of $q$.  This agrees with the Eulerian form presented by \citealt{nof1990} and \citealt{Siegelman2022}.}

\bibliographystyle{jfm}
\bibliography{ref}

\end{document}